\documentclass[a4paper,twocolumn,10pt]{article}
\usepackage[margin=0.7in]{geometry}

\usepackage{todonotes}
\usepackage{fancyhdr}

\begin{document}
	
	\title{\centering“Cyber security is a dark art”: The CISO as soothsayer}
	
			\author{}
	{
		\author{
						{\rm Joseph Da Silva}\\
			Royal Holloway, University of London\\
			joseph.dasilva.2018@live.rhul.ac.uk
			\and
			{\rm Rikke Bjerg Jensen}\\
			Royal Holloway, University of London\\
			rikke.jensen@rhul.ac.uk
		}
		
	}

\maketitle

\begin{abstract}
		Commercial organisations continue to face a growing and evolving threat of data breaches and system compromises, making their cyber-security function critically important. Many organisations employ a Chief Information Security Officer (CISO) to lead such a function. We conducted in-depth, semi-structured interviews with 15 CISOs and six senior organisational leaders, between October 2019 and July 2020, as part of a wider exploration into the purpose of CISOs and cyber-security functions. In this paper, we employ broader security scholarship related to ontological security and sociological notions of identity work to provide an interpretative analysis of the CISO role in organisations. Research findings reveal that cyber security is an expert system that positions the CISO as an interpreter of something that is mystical, unknown and fearful to the uninitiated. They show how the fearful nature of cyber security contributes to it being considered an ontological threat by the organisation, while responding to that threat contributes to the organisation's overall identity. We further show how cyber security is analogous to a belief system and how one of the roles of the CISO is akin to that of a modern-day soothsayer for senior management; that this role is precarious and, at the same time, superior, leading to alienation within the organisation. Our study also highlights that the CISO identity of protector-from-threat, linked to the precarious position, motivates self-serving actions that we term `cyber sophistry'. We conclude by outlining a series of implications for both organisations and CISOs. 
\end{abstract}

\section{Introduction}\label{sec:introduction}
An effective cyber-security function is of critical importance to commercial organisations, particularly due to the continuing and very public threat of data breaches and system compromises. As threats evolve in terms of sophistication from both external and internal sources, having an in-house capability that holds the responsibility for effective management of the organisation's information security posture is a common means of risk mitigation. In order to lead such a capability, many organisations employ a Chief Information Security Officer (CISO). We argue that one of the roles the CISO performs is akin to that of a modern-day soothsayer, grounded in an understanding of cyber security being reliant on specialised and sometimes mystical insights and skills. We situate our analysis within broader security scholarship related to ontological security and sociological notions of identity work and, through such analytical lenses, deepen the understanding of the CISO role within organisations. Our findings show that CISOs sit at a nexus, assimilating information from multiple sources, making judgements and relaying those judgements to senior management in terms those stakeholders can understand. The analogy of soothsaying, a term we neither consider nor intend to be denigratory, acts as both a rhetorical device that brings to light the difficult and conflicting position that CISOs occupy and as a model with which to explore their role.

The motivation behind our study is to explore and understand the purpose of CISOs within commercial businesses. CISOs perform an important, challenging, and yet poorly understood and ill-defined role~\cite{hooperEmergingRoleCISO2016,ashendenCISOsOrganisationalCulture2013}. By focusing on perceived purpose, both from the perspective of CISOs and from the perspective of their most senior stakeholders, across a relatively large and varied sample, we aim to gain a greater understanding of CISO practice that offers value to practitioners and their employers -- and to CSCW and cyber-security scholarship. To that end, we do not explore how purpose may be perceived at other organisational levels or by other stakeholders. While we focus on the soothsayer identity in this paper, as it emerged through our analysis, it is important to note that it is one of a number of different roles we identified CISOs performing. Future work will explore other roles, including the CISO as police and the CISO as salesperson.

\paragraph{Contributions.}\label{sec:contributions}
Our work contributes to and extends existing, albeit limited, scholarship which has explored cyber security within organisations, e.g.~\cite{CSCW:DKDH15,CSCW:GoLuKo04,CSCW:JWXRC15,CSCW:PKTEK17,kockschCaringITSecurity2018,singhInformationSecurityManagement2013,werlingerSecurityPractitionersContext2009}, and an even smaller number focusing on CISOs, e.g.,~\cite{ashendenCISOsOrganisationalCulture2013,reinfelderSecurityManagersAre2019}. We do so by studying the role and identity of the CISO as critical to the organisational cyber-security function and by bringing it into conversation with wider security scholarship related to ontological security and sociological notions of identity work, through a theoretically grounded interpretative analysis.

Data was gathered through 21 semi-structured interviews with CISOs (15) and senior organisational leaders (six) representing a range of commercial organisations, predominantly multinationals based in the United Kingdom (UK). The interviews were carried out between October 2019 and July 2020. The data was analysed thematically, following Braun and Clarke~\cite{braunUsingThematicAnalysis2006,braunReflectingReflexiveThematic2019}, with analytical techniques influenced by Salda\~{n}a~\cite{saldanaCodingManualQualitative2016}. One key theme that emerged from the data is that the CISO acts as a soothsayer to the organisation, both in their own approach and in the expectations of their stakeholders. We identify different forms of identity work performed by the CISO that support the CISO-as-soothsayer construction and link this to wider security themes.\footnote{While the concept of soothsaying has been applied to other professions such as law, e.g.~\cite{osbeckLawyerSoothsayerExploring2018,johnsonSoothsayersLegalCulture2020}, no previous work, to our knowledge, has substantively made the connection to cyber security.}

To this end, our work contributes four key findings, which are outlined in Section~\ref{sec:findings} and discussed in Section~\ref{sec:discussion}. First, our study shows how cyber security is an expert system that both demands interpretation by someone with specialised knowledge and is opaque to non-specialists. This positions the CISO as an interpreter to key stakeholders. Second, cyber security appears mystical, unknown and fearful to the uninitiated; the non-specialists. We observe how this makes cyber security analogous to a belief system, which positions the CISO-as-interpreter as a soothsayer. Third, as a soothsayer, the CISO occupies a precarious role, at a number of levels. While they are often scapegoated when things go wrong, their `superior' position of special knowledge also means that they often assume a position of an outsider -- or `other' -- in the organisation. Fourth, cyber security is implicated in identity, for both CISOs and the organisations that they work for. Our study shows how CISOs as well as organisations experience ontological insecurity, which leads to the development of identities as protector-against-threat and protected-from-threat, respectively. For CISOs, we show how this identity motivates self-serving actions that we term `cyber sophistry'.

We conclude by setting out a number of implications that the construction of CISO-as-soothsayer has for organisations. Moreover, we consider how a broader understanding of and approach to cyber-security research can benefit future research in CSCW and adjacent fields as well as the role of the CISO.

\section{Related work}\label{sec:related-work}
In this section, we position our work within existing scholarship on cyber security in organisational contexts and in relation to wider CSCW work on cyber security, focusing on work that is both psychologically and sociologically driven, while bridging to notions of identity work, ontological security and permanent emergency. 

\subsection{Cyber security in organisations}
Prior research into cyber-security practice in organisations, much of which has also utilised semi-structured interviews~\cite{ashendenCISOsOrganisationalCulture2013,singhInformationSecurityManagement2013,werlingerSecurityPractitionersContext2009}, has identified a lack of understanding and confusion regarding the subject~\cite{haneyItScaryIt2018}. Security improvements may be seen as of little value within organisations~\cite{pollerCanSecurityBecome2017}, with security itself being seen as ``scary \dots{} confusing \dots{} dull''~\cite[p. 411]{haneyItScaryIt2018} or focused on attributing blame for failings~\cite{kockschCaringITSecurity2018}. A number of studies have investigated cyber-security behaviours within organisations, e.g.~\cite{nicholsonIntroducingCybersurvivalTask2018, beautementProductiveSecurityScalable2016,conwayQualitativeInvestigationBank2017}, and the effectiveness of programmes aimed at improving such behaviours, e.g.~\cite{reinheimerInvestigationPhishingAwareness2020, siadatiMeasuringEffectivenessEmbedded2017}, with some concluding that top-down, policy-based approaches to improving cyber-security behaviour were ineffective~\cite{kirlapposComplyDeadLong2013} and others finding more effective results from game-based interventions~\cite{jayakrishnanPassworldSeriousGame2020}. Poller et al.~\cite{CSCW:PKTEK17} call for security initiatives implemented in organisations to be ``reconciling people's ideas of security with the requirements of the organizational setting''~\cite[p.2502]{CSCW:PKTEK17}; thus, highlighting the centrality of both communication and management perspectives. 

\subsubsection{CISOs}
Cyber security may be perceived as an individualistic practice and thus not `cooperative' as noted by Goodall et al.~\cite{CSCW:GoLuKo04}. Their study showed that cyber security is inherently collaborative both within and across organisational settings -- while also being grounded in a community of experts. Effective cyber security depends on multiple actors, not just those who are cyber-security practitioners~\cite{reinfelderSecurityManagersAre2019}. However, a number of studies highlight that CISOs appear be somewhat disconnected from the rest of their organisations, being seen as blockers~\cite{ashendenCISOsOrganisationalCulture2013}, governors~\cite{reinfelderSecurityManagersAre2019,karanjaRoleChiefInformation2017}, translators~\cite{hooperEmergingRoleCISO2016}, or even adversaries~\cite{ashendenSecurityDialoguesBuilding2016}. They also highlight a lack of clarity regarding what a CISO is expected to do~\cite{hooperEmergingRoleCISO2016,karanjaRoleChiefInformation2017}, and how CISOs experience conflicted identities~\cite{ashendenCISOsOrganisationalCulture2013}. CISOs may be recruited in response to a cyber-security exposure~\cite{karanjaRoleChiefInformation2017}, and thus occupy an identity that ‘fixes’ rather than ‘prevents’, yet may implement controls that are negatively perceived~\cite{reinfelderSecurityManagersAre2019} or ignored~\cite{kirlapposComplyDeadLong2013}. They may be distrusted by their business stakeholders~\cite{ashendenSecurityDialoguesBuilding2016} and find that there is a broader lack of understanding as to what they actually do~\cite{zanuttoShadowWarriorsNo2017}.

While these other works have problematised cyber security in organisational contexts, and highlighted confusion in relation to the role of the CISO, our work provides a different perspective. Beyond highlighting murkiness regarding the role, e.g, ~\cite{ashendenCISOsOrganisationalCulture2013,hooperEmergingRoleCISO2016,karanjaRoleChiefInformation2017}, and the existence of disconnects between security managers and the rest of the organisation~\cite{reinfelderSecurityManagersAre2019}, we present a discussion that explores one of the multiple roles of a CISO and the implications this has on cyber-security practice in business, employing theoretical concepts as analytical lenses to deepen interpretations of empirical data. The latter is not a new phenomenon in CSCW work, albeit one that is significantly limited and under-explored, despite the fact that the benefits of (organisational) theory for the field was discussed at a 2004 CSCW panel~\cite{CSCW:BDKRKY04}. Here, some of the panellists argued for more theoretically informed approaches to CSCW research, including theories from sociology and social psychology. Echoing Kocksch et al.~\cite{kockschCaringITSecurity2018}, who used the notion of care as an analytical lens to bring to light the often invisible aspects and practices that underpin IT security, we contend that an approach that employs analytical lenses grounded in organisational and sociological concepts is of significant value to CSCW and HCI research more broadly, both with regards to academic scholarship and practitioner practice. We also base our study on a more extensive sample of participants than many other studies of CISOs, e.g.,~\cite{ashendenCISOsOrganisationalCulture2013,reinfelderSecurityManagersAre2019} as well as providing a broader perspective derived from a sample of senior stakeholders.

\section{Theoretical grounding}\label{sec:theory}
In this section, we provide a brief overview of theories of identity work, ontological security and permanent emergency. We employ these theoretical positions to interpret and produce meaning from the research findings, outlined in Section~\ref{sec:findings}.

\subsection{Identity work}
Identity work is an established concept in the social sciences, with the term originating from a seminal work by sociologists Snow and Anderson~\cite{snowIdentityWorkHomeless1987}, but drawing significantly on earlier work by Goffman, e.g.~\cite{goffmanAsylumsEssaysSocial2017}. In essence, identity work refers to the multiple discursive strategies employed by individuals in constructing and maintaining their identity. Ybema et al summarise this succinctly by “suggest[ing] that ‘identity’ is a matter of claims, not character; persona, not personality; and presentation, not self” \cite[p. 306]{ybemaArticulatingIdentities2009}. Others have applied and expanded this concept, including its application to organisational research, e.g.~\cite{brownIdentitiesIdentityWork2015,cazaIdentityWorkOrganizations2018,knappManagingBoundariesIdentity2013}. Identity work can include the construction of identities that are deemed to be `moral', with those who are ``known for exemplary moral commitments'' referred to as ``moral exemplars''~\cite[p. 497]{hardyMoralIdentity2011}.\footnote{While the existence of `moral experts' is debatable~\cite{driverMoralExpertiseJudgement2013}, the concept is well established and continues to be applied, e.g.~\cite{riazHowIdentifyMoral2020}.} Morality has also been linked with cyber security, e.g.~\cite{nissenbaumWhereComputerSecurity2005,kockschCaringITSecurity2018}. 

Identity may be threatened, a concept explored by several scholars, e.g.~\cite{ellemersSelfSocialIdentity2002,kreinerWhereMeWe2006,beechIdentityintheworkMusiciansStruggles2016}. Although well established~\cite{brownIdentityThreatsIdentity2015}, definitions of `identity threat' are contested~\cite{petriglieriTHREATRESPONSESCONSEQUENCES2011}. For Petriglieri, threats to identity have the ``unique feature of arising from present cues of future harm''~\cite[p. 644]{petriglieriTHREATRESPONSESCONSEQUENCES2011}. This can include unrealistic or unwanted role expectations~\cite{ibarraIdentityWorkPlay2010}. Identity threats are considered to motivate identity work~\cite{ibarraIdentityWorkPlay2010,brownIdentityThreatsIdentity2015}, in order to maintain the (present) identity and by extension, ontological security (we expand on this in Section~\ref{sec:ontological-security}). Cunliffe suggests that, through dialogic identity construction, individuals build a sense of ontological security~\cite{cunliffeManagersPracticalAuthors2001}.  

Identity work has been considered in relation to an organisation's overall identity, e.g.~\cite{kreinerElasticityDialecticTensions2015} and ``[o]rganizational legitimacy is also a key [identity] concern, because legitimate status is a \textit{sine qua non} for easy access to resources, unrestricted access to markets, and long-term survival''~\cite[p. 38]{brownNarrativePoliticsLegitimacy1998} (italics in original). The notion of identity threat has also been applied to collectives, e.g.~\cite{ellemersSelfSocialIdentity2002,brownIdentityThreatsIdentity2015}.

\subsection{Ontological security}\label{sec:ontological-security}
Ontological security, as a concept, has been explored in some detail within the fields of International Relations and Critical Security Studies, e.g.~\cite{huysmansSecurityWhatYou1998,roeValuePositiveSecurity2008,mcsweeneySecurityIdentityInterests1999}, as well as by those in Sociology~\cite{giddensConsequencesModernity1990}, Psychiatry~\cite{laingDividedSelfExistential2010} (where the concept arguably originated), and a number of other disciplines including cyber security, e.g.~\cite{coles-kempNewLandMobile2018} and Organisational Studies, e.g.~\cite{thompsonPeoplePracticeTechnology2012}.

Steele defines ontological security as ``security as being'' rather than ``security as survival''~\cite[p. 426]{steeleOntologicalSecurityPower2005}. This aligns with Giddens~\cite{giddensConsequencesModernity1990} but contrasts with Buzan~\cite{buzanPeopleStatesFear1991} and Huysmans~\cite{huysmansSecurityWhatYou1998}. For Buzan, survival is ``the bottom line''~\cite[p. 19]{buzanPeopleStatesFear1991}. Coles-Kemp et al. summarise McSweeney's~\cite{mcsweeneySecurityIdentityInterests1999} position on the term as both ``the freedom to live free from fear as well as protection from harms''~\cite[p. 3]{coles-kempNewLandMobile2018}, echoing the thoughts of Roe~\cite{roeValuePositiveSecurity2008}.

McSweeney discusses the \textit{need} for insecurity and how it has underpinned international relations ``for centuries''~\cite[p. 2]{mcsweeneySecurityIdentityInterests1999}. Burke~\cite{burkeAporiasSecurity2002} describes security's dependence on insecurity; how it is defined by it. This insecurity is often achieved through the maintenance of a state of `permanent emergency', which may be used to achieve control and set boundaries within a society~\cite{neocleousCritiqueSecurity2008,bubandtVernacularSecurityPolitics2005}. This concept has been debated for many years~\cite{grossNormlessExceptionlessException2000}, with Neocleous~\cite{neocleousCritiqueSecurity2008} providing historical examples, to demonstrate how `temporary' emergency powers have been used to limit workers' rights and regulate behaviour~\cite{neocleousProblemNormalityTaking2006}, and how such powers have subsequently become permanent, including those that erode civil liberties~\cite{tunnellPissingDemandWorkplace2004}. 

Neocleous further notes how the continued existence of threats and an associated atmosphere of fear is needed to justify security-related actions, including the necessary existence of ``demons \dots{} villains''~\cite[pp. 119, 223]{neocleousCritiqueSecurity2008}. Juntunen and Virta describe the potential for security crises to be viewed as ``almost \dots{} desirable''~\cite[p. 72]{juntunenSecurityDynamicsMultilayered2019} in the sense of building resilience among those affected and describing the ``normalis[ation of] crises \dots{} as potentially positive learning experiences''~\cite[p. 80]{juntunenSecurityDynamicsMultilayered2019}.

\section{Methodology}\label{sec:methods}
In this section, we set out our methodology, including the methods used, the recruitment and interview process, ethical considerations, data analysis and limitations.

\subsection{Semi-structured interviews}\label{sec:methods-interviews}
Data was collected through semi-structured interviews and these were all conducted by one of the researchers, who happens to be a practising CISO. Interviews took place between October 2019 and July 2020. In total, 21 interviews were conducted, within 18 different companies. Of the 21 interviews, 15 were with CISOs and the remaining six were with senior organisational leaders (see Table~\ref{tab:interviews}). We refer to participants by their abbreviated job role and an incremental number, i.e., CISO1, CISO2, CEO1. Interview guides were prepared for each set of participants, i.e.~CISOs and non-CISOs, comprising a broad set of question topics to use as prompts when required. We include these in Appendices~\ref{sec:topic-guide-ciso} and~\ref{sec:topic-guide-nonciso}. The interview design was based on the primary research question: \emph{what is the purpose of a CISO within a commercial organisation?}


\begin{table}[h!]
\centering
	\footnotesize
	\caption{Participants \& Interviews}\label{tab:interviews}
	\begin{tabular}{ll @{\hskip 2em} lll}
		\multicolumn{2}{c}{\textbf{Participants}} & \multicolumn{3}{c}{\textbf{Interview}}\\
		\emph{ID} &  \emph{Duration} & \emph{Medium} & \emph{Timing} \\
		CISO1 & 00:48:29 & F2F & Oct19 \\
		CISO2 & 00:49:28 & F2F & Oct19 \\
		CISO3 & 00:47:33 & F2F & Dec19 \\
		CISO4 & 00:44:41 & F2F & Dec19 \\
		CISO5 & 00:43:44 & F2F & Dec19 \\
		CISO6 & 00:41:38 & F2F & Jan20 \\
		CISO7 & 00:45:19 & F2F & Jan20 \\
		CISO8 & 00:49:41 & F2F & Mar20 \\
		CISO9 & 00:51:30 & F2F & Mar20 \\
		CISO10 & 00:38:43 & Remote & Apr20 \\
		CISO11 & 00:55:45 & Remote & May20 \\
		CISO12 & 00:40:56 & Remote & May20 \\
		CISO13 & 00:40:07 & Remote & Jun20 \\
		CISO14 & 00:46:07 & Remote & Jul20 \\
		CISO15 & 00:50:02 & Remote & Jul20 \\
		CEO1 & 00:24:59 & F2F & Dec19 \\
		CEO2 & 00:42:45 & F2F & Jan20 \\
		CFO1 & 00:45:41 & F2F & Jan20 \\
		CFO2 & 00:40:52 & Remote & Apr20 \\
		CIO1 & 00:47:28 & Remote & Jul20 \\
		NED1 & 00:27:52 & F2F & Dec19 \\
	\end{tabular}
	
	\vspace{1em}
	Note: in addition to 15 CISOs, we conducted semi-structured interviews with two Chief Executive Officers (CEOs), two Chief Financial Officers (CIOs), one Non-Executive Director (NED) and one Chief Information Officer (CIO), between October 2019 and July 2020. 
\end{table}

\subsubsection{Interview process} 
Interviews were carried out either face-to-face or online. For face-to-face interviews, participants were interviewed at their own office locations. Being in their `own space' provided participants with a sense of security, both physical and psychological. Interviews conducted online were done so in response to the COVID-19 pandemic which occurred during the data collection phase. Interviews were recorded and transcribed as soon as possible following each interview. Significant aspects of the interview, not captured in audio recordings, such as body language and spatial information, were captured in a handwritten journal immediately following each interview.

As mentioned above, interview guides were prepared in advance. These were not intended to be an exhaustive list of questions, rather they were example prompts that the researcher used to guide the conversation as and when appropriate. Interviews were approached as a conversation focused on cyber-security practice, which was a joint construction between the researcher and the participant, rather than as an extraction of data~\cite{alvessonNeopositivistsRomanticsLocalists2003}. Interviews were scheduled for one hour in total, however, some conversations ended naturally before the end of the time allocated, and some participants had to cut the interview short due to unplanned demands on their time that emerged on the day.

\subsubsection{Participants and recruitment} 
Participants were predominantly CISOs, with a smaller pool of executives and non-executives, from a variety of UK-based, but predominantly multinational, commercial entities. Organisations were selected primarily on the basis of access through professional networks. However, one criterion applied was that research was limited to organisations that were quoted on a major stock market, which was intended to provide a level of consistency across organisations on the basis of them being subject to consistent corporate governance requirements. 

We used our own networks of professional contacts to recruit participants. This meant that the eventual sample was effectively a ``snowball sample''~\cite[p. 135]{hammersleyEthnographyPrinciplesPractice1995} with a number of contacts introducing additional participants. This occurred both through those who were directly involved in the research and were able to introduce others as well as through those who were not involved in the research. The latter group included acquaintances at other companies who did not work in cyber security or at a board level, but were willing to make introductions to relevant contacts. In general terms, each contact led to the recruitment of one participant. This helped to ensure that our sample was as diverse as possible. There is little opportunity to gain access to board members outside of a professional environment and access to these participants was approached through CISOs who were participating in the research, as well as through board members with whom we already had a relationship. Although the intention was to gain both CISO and senior leader perspectives from each organisation, due to the difficulty in gaining access, only six non-CISO participants were recruited (see also Section~\ref{sec:limitations}). Participants were engaged through a variety of means, including direct email contact, face-to-face interaction at industry events, the use of mutual acquaintances to effect introductions and direct contact on LinkedIn. The latter method was the least successful but did result in a small number of interviews.

\begin{table}[h!]
\centering
  \footnotesize
  \caption{Industry sectors represented in this study}\label{tab:industries}
  \begin{tabular}{ll @{\hskip 2em} lll}
    \emph{ICB Super-sector} & \emph{Number of organisations} \\

Banks & 1 \\
Food, Beverage and Tobacco  & 1 \\
Industrial Goods and Services & 6 \\
Personal Care, Drug and Grocery Stores & 2 \\
Real Estate & 1 \\
Technology & 1 \\
Telecommunications & 2 \\
Travel and Leisure & 1 \\
Utilities & 3 \\
 \textbf{Total} & \textbf{18}
     \end{tabular}
  
  \vspace{1em}
Coverage of industries represented in this research based on classifications taken from~\cite{IndustryClassificationBenchmark2018,CompaniesSecuritiesLondon}.
\end{table}

\subsubsection{Ethical considerations}
Every care was taken to protect participants at every stage of the research. All of our activities were approved for self-certification through our institution's Research Ethics Committee, before the start of the research. The study was designed to minimise both the collection of personally identifiable information and the risk of indirect identification. Participants were fully informed of the research and were provided with consent forms and information sheets two working days before each interview; the information sheet explained how data would be anonymised and protected. We include the information sheet in Appendix~\ref{sec:pis}. All interview data was anonymised. We only refer to participant IDs, their role, the company that they represented and the industry sector applicable to that company. Participants were referred to only by their ID from the point of transcription onwards. Any sensitive or potentially identifiable information was redacted during transcription. Once interviews were transcribed and anonymised all recordings were destroyed.

\subsection{Data analysis}\label{sec:analysis}
The data was collected as part of a wider study exploring the purpose of cyber-security functions within commercial organisations. The findings presented in this paper represent one theme emerging from this work, namely the conception of the CISO as a soothsayer. This notion manifested itself across the data. Transcripts were analysed inductively and coded in multiple cycles using NVivo 12~\cite{QualitativeDataAnalysis}. We used a mixture of coding types, with in-vivo, process, versus and dramaturgical coding~\cite{saldanaCodingManualQualitative2016} being used extensively. Following the inductive phases, we applied a deductive approach in order to categorise and rationalise the codes. Analytic memos were produced throughout, following others, e.g.~\cite{saldanaCodingManualQualitative2016,kreinerTabulaGeminusBoth2016,charmazGroundedTheoryMethodology2015}. Diagramming was used extensively in the analysis (see Appendix~\ref{sec:diagram} for an example) to explore relationships between codes and categories, combining several methods from Salda\~{n}a~\cite{saldanaCodingManualQualitative2016} including category relationship and operational model diagrams. These were used to identify and explore themes emerging from the data, following~\cite{braunUsingThematicAnalysis2006}. At this stage, multiple sociological lenses were used to facilitate a deeper interpretation and sense-making of our research findings.

\subsubsection{Limitations}\label{sec:limitations}
A number of limitations should be taken into account when interpreting our findings. First, while a broad range of industries are represented in the study, there is a weighting towards one industry (Industrial Goods and Services) and there are industries that are not represented. Second, the semi-structured nature of the interviews was chosen to provide depth rather than scale, and the findings will not be representative of all CISOs or senior leaders, nor of all companies. Third, there is an inherent bias in interview-based research, given that participants self-select to take part. It may be the case that the participants who agreed to be interviewed were those that were particularly motivated to express identity concerns. This limitation is not unique to this study, but mirrors other interview-based studies. Fourth, there is a lack of diversity in the study, at a number of levels. The majority of the participants, and all of those in CISO roles, have self-identified as male. All participants can also be considered to be `elites', while there is a lack of ethnic diversity in the study. This lack of diversity, however, reflects the lack of diversity in the industry more broadly. Fifth, the positionality of the main researcher, as a practising CISO, may be a crucial factor in the effectiveness of the analysis stage~\cite[p. 8]{caelliClearMudGreater2003}. It may have aided interpretation, although with the risk of bias. Sixth, there is an imbalance between CISO and non-CISO participants, due to the difficulty in securing both engagement with, and commitment from, senior leaders. Seventh, other organisational levels beyond senior leaders are not reflected in this study and, therefore, roles that the CISO plays at these other levels has not been explored. Eighth, the interviewer was a CISO, which will have introduced aspects of personal bias. 
Finally, the COVID-19 outbreak meant that eight of the interviews were conducted remotely (see Table~\ref{tab:interviews}). Conducting interviews online limited the researcher's ability to observe the participants' physical setting, which might have affected their ability to speak freely.

\section{Research findings}\label{sec:findings} 
Our research findings are structured into five sections. Section~\ref{sec:interpretative-practice} focuses on the role of the CISO as cyber-security interpreter and communicator, Section~\ref{sec:expert-system} covers cyber-security expertise and associated trust, Section~\ref{sec:mystical} highlights perceived mystical aspects of cyber security, Section~\ref{sec:fearful} focuses on the relationship between fear and cyber security, and Section~\ref{sec:precarity} shows the CISO role as precarious and alienated.

\subsection{Cyber security as an interpretive practice}\label{sec:interpretative-practice}
Multiple participants described there being a \textit{need} for cyber-security decisions to be made based on interpretation. These did not solely originate from CISO participants. For example, CFO2 described how cyber-security information needs to be ``expressed in a way that is understandable to a more lay person''. This indicates both a sense at the senior leadership level that such information needs to be interpreted, or at least translated, but also that such information is specialist, perhaps even mystical. Although the term ``lay person'' may be used to indicate a non-specialist, it carries implications of secular versus clerical and may, therefore, also suggest a mystical nature to cyber security. CISO14 suggested that they met the needs of CFO2 when they stated that:

\begin{quote}
 ``I don't just say that there's this flavour of ransomware and it's really bad and it could hit us, that means nothing, you know, I try to articulate that in a balanced risk-based way so that they can understand how vulnerable we are \dots{} what the consequence would be perhaps in financial terms, yeah, and what we need to do and how much it might cost to mitigate that risk.'' 
\end{quote}

\noindent Not only have they interpreted the information from the perspective of its relevance to the organisation and its exposure, they have done so, according to them, in a balanced way, using language that they consider to be more understandable by their audience. This suggests that the CISO is an interpreter and translator, and holds a position of power as a gatekeeper of information. 

\subsubsection{Cyber security as a foreign language}
As well as information needing to be interpreted, it needs to be relayed in words that the audience can understand. CISO4 described how they ``speak in a language that actually is very business-specific rather than technology-specific'', they ``keep it as simple as we can \dots{} as monosyllabic as we can''. CISO8 noted how they wanted to ``find a common language for talking to the board about security''; this common language was financial in nature, similar to the explanation given by CISO14 above. 

CISO5 made an explicit reference to cyber security being a foreign language, describing how ``when you try to talk to anybody in another language you should at least know how to say please, thank you, order a beer'' and stating that it is ``incumbent upon a CISO or any head of security in whatever guise to teach the recipient how to speak pidgin cyber or pidgin IT or something that gives them a fighting chance to understand''. The same participant described how ``we almost speak in you know, hieroglyphics or we speak in \dots{} language that nobody else speaks in''. The reference to hieroglyphics implies some mysticism regarding the language of cyber security, which is discussed further in Section~\ref{sec:mystical}.

Several references were made to cyber security being difficult to understand. Multiple CISOs described the necessity of analogy and metaphor in their communications, particularly with senior leaders. The CISOs also indicated that their stakeholders had a limited understanding of the subject, across all levels of the organisation. Senior leaders expressed the need for cyber-security messaging to be kept ``as simple as possible'' (CEO2) and to be ``expressed in a way that is understandable to a more lay person'' (CFO2). Moreover, CFO2 described how cyber-security specialists ``have a depth of knowledge and a smell and a sense for their particular area''. This also suggests something beyond knowledge, an intangible, arguably mystical, ``sense'' for the subject. 

\subsubsection{Cyber security as a continuum}
Participants indicated that cyber security represents a continuum, with the implication being that, due to the lack of a binary nature, interpretation is necessary. CFO2 highlighted the interpretive aspect of cyber security by stating that ``we don’t gold-plate it I wouldn't have said, but we don't short-change it'', implying that there is a `range', a continuum from short-changing to gold-plating, with regard to security. This also implies that it is necessary to determine where on this scale an organisation should be placed. CISO7 used similar language when describing how they agreed risk tolerance with their board as ``we're not going to be gold-plated''. They further implied a continuum by distinguishing between ``good practice'' and ``best practice'' and added that not achieving best practice amounted to ``just common sense stuff''.\footnote{This perhaps undermines the idea of there being a need for a specialist interpreter. However, it is likely the case that this is only `common' to specialists.} They further noted a ``subtlety between `do you want good practice or best practice'\,'' when talking to the board about risk tolerance. By asking the board whether they want good practice or best practice, they are both constructing the continuum and their role as an interpreter of the continuum, and of being capable of helping their stakeholders determine where on that continuum they `should' be. They admitted to ``the subjectiveness of it'' and described part of their role as ``wordsmithing things'', further indicating the `art' of what they do but also the opportunity for sophistry.

Other references to a continuum were made by participants, including the role that the CISO plays in determining where on that continuum the organisation should be. For example, CISO12 described ``a certain baseline they need to achieve and in security that's where I come in'' and how, within their organisation, they set each individual business unit a ``bar'' to achieve with regard to cyber security. Moreover, several CISOs invoked a `one size doesn't fit all' argument with reference to cyber security, implying the need for a `tailored' approach. This further implied the need for a specialist, a `tailor', who will interpret the needs of the situation and design an appropriate solution, with the CISO positioning themselves as this tailor, this interpreter. CISO4 described how ``it's quite complex \dots{} there's no one size \dots{} that we prescribe'', invoking not just the language of a tailor but that of a medical practitioner as well. There is also an implication here of power; the CISO is in a position to be able to prescribe to the organisation what its approach to cyber security should be. 

\subsubsection{Cyber security as not-science}
Several CISOs either explicitly labelled themselves as pragmatic or as not-dogmatic, or implicitly invoked that label using similar words, e.g. ``I think they appreciate I'm pretty candid and also realistic'' (CISO2). The invocation of `pragmatic' (identified in 13 out of 15 CISO interviews) implies a sense of interpretation; in order to be pragmatic or realistic, both an analysis and a judgement are required. 

In addition, several CISOs described cyber security as not-science; CISO2 passionately stated that ``although much of \dots{} cyber security \dots{} is kind of positioned as almost a science, you know my own approach is that it's not \textbf{at all}'' (emphasis captured in original transcript). They added that they rely on ``judgement and gut feeling and stuff like that'', suggesting a self-conception as a soothsayer, and also as a judge. They further highlighted how they determine ``the right course of action'' based on both ``data'' and ``gut feeling'', but that often there is no empirical data and therefore they must rely on their own ``experience''. 

In line with this, several CISOs invoked the `art not science' motif by describing the immeasurability of the cyber-security risk and pressures from their stakeholders for measurability and quantification. This resulted in the CISOs feeling frustrated, as they considered that their stakeholders did not appreciate that ``there isn't a clear answer'' (CISO13).

\subsection{Cyber security as an expert system}\label{sec:expert-system}
CFO2 described how ``the cyber team recommend to us as a management team what they think they need to do because it's quite specialised'', which may suggest that this ``cyber team'' position themselves as specialists. Indeed, making recommendations is one method by which an identity as an expert may be constructed. This participant also referred to how they need to ``unpick'' the information they are provided, in which ``there are quite a lot of acronyms''. CISO3 described how their board ``still struggle a little'' and explicitly positioned themselves as an expert, describing how ``you'll come in [to a board meeting] as a subject matter expert''. Through this narrative construction, they are not only the expert but also the solution to a problem. Another CISO suggested a similar narrative: ``when I turned up, it was management by anecdote \dots{} the board didn’t know which question to ask \dots{} I think I've turned that around''. CISO4 described ``dealing with \dots{} a particularly sophisticated set of attack activity'', adding that ``cyber-security is a dark art \dots{} most people won't understand it''; their narrative describes a problem that is both fearful and incomprehensible.

CIO1 described the need for any individual who is responsible for cyber security to have ``the credentials in place to do that first'' and describes how their staff have ``incentives to gain qualifications quickly''; these references suggest a perceived need not just for experts but for proof of that expertise. Other indications of constructing cyber security as an expert system include references to the subject itself being demanding; CISO11 for example stated how ``it's not simple and it's not easy, it's not straightforward''. CISO1 alluded to this when describing how they ``want to give them [their senior stakeholders] an idea of the scope of what we're trying to do''. By giving stakeholders the idea that the CISO is dealing with something that is difficult, they also construct (or maintain) an identity as an expert.

\subsubsection{Trust in expert systems}
The non-CISO participants considered cyber security to be an expert system. CFO2, for example, referred to how their CISO is ``bringing real time expertise to the organisation''. NED1 described how ``we're [i.e. board members] not going to have the deep expertise'' and CFO1 stated that they are ``never going to be a cyber-security expert'', how cyber security itself is ``a topic \dots{} where you're always going to be looking to genuine experts''. Not only is cyber security accepted as an expert system, it is trusted. NED1 described how ``we're going to trust that we're not going to have the deep expertise \dots{} but we do ultimately have to understand the right questions to ask to make sure someone like you [i.e. the interviewer, a CISO] is doing your job appropriately.'' CFO1 noted that they would ``rather someone \dots{} would look at it [cyber security] properly and, you know, knew what it should look like''. CISO6 described the importance of trust, but suggested the possibility of abusing that trust:

\begin{quote}
 ``I like to make sure that we are very honest about the risks because it's very easy to mislead because \dots{} they [the board] don't know enough \dots{} so they have to have some faith and trust in me and obviously my boss for representing that.'' 
 \end{quote}

Several participants, both CISOs and non-CISOs, referred to external standards such as NIST-CSF, ISO27001 and ISO31000. For example, CFO2 described how they would ``like us to have a more mature ISO 27001 environment''. As well as the incantation of standards, participants also referred to accreditation against them. This provides a further means of cementing cyber security as an expert system; not only is there an external standard, agreed upon by external experts and branded with/by a recognised external authority, compliance with that standard can be assessed by further experts, i.e. auditors. CISO9, however, partly challenged the use of standards, saying that ``ISO27001 is often seen as a bit of a panacea'' and that they are ``not necessarily convinced that ISO[27001] is the sort of \dots{} big sticking plaster that everybody thinks it is''. They suggested that there is a ranking of accredited standards, that some are considered ``lesser certifications or qualifications'', but they think that ``actually \dots{} they all have a place''. Moreover, CISO7 described how their board ``quite like the fact that I've based us around the NIST cyber-security framework'' and described referencing this standard in their reporting, using it to frame their activity. The use and presence of standards may serve to legitimise their activity but also their position as an interpreter and, by extension, an expert. One CISO suggested that if a provider of a service is regulated, that fact ``slightly obviates'' the consumers of that service ``from having to make a judgment''. Similarly, the reference to a standard by a CISO may ``obviate'', at least in part, the need of their audience to make a judgement on the CISO's legitimacy or expertise. 

\subsubsection{Expert superiority}
A further suggestion of an expert identity was made by CISO3 who stated ``it's like children isn't it'', when describing how they need to communicate cyber risk to their stakeholders; they described how ``people don't like to be told what to do \dots{} unless they understand''. This suggests not just the construction of an expert identity but also one of superiority, a more knowledgeable and well-meaning parent explaining something to a less knowledgeable, naïve child. A different aspect of superiority was expressed by CISO14 when referring to one of their team members; they described how ``you might not put him in front of the board of directors''. As well as being superior, this may suggest that a particular `type' of person is required in order to interact with senior leaders.\footnote{A related finding from this study was a clear differentiation between technical and non-technical identities; this will be explored in future research.} CISO4 stated that ``most people won't understand it [cyber security]''; this implies not just the specialist nature of the subject, but also the specialist nature (in the sense of \textit{being} `special') of the person who does understand it. By delineating between those who do and those who do not understand there is also a suggestion of superiority, that having that understanding sets them apart from ``most people''. 

\subsection{Cyber security as mystical or arcane}\label{sec:mystical}
Not only is the subject complex, it is ``a dark art'' (CISO4), it is ``another language'' (CISO5), ``almost \dots{} hieroglyphics'' (CISO5). Those who do not practice it are ``lay-person[s]'' (CISO2) and those outside of it, such as CFO2, recognise themselves as such. The use of this language helps to position cyber security as a specialist subject, but also as something mystical, possibly even clerical in nature. As a subject that is specialist and mysterious, it requires a specialist to make sense of it. If positioning cyber security as mystical benefits the conception of it as an interpretive practice, which therefore requires an interpreter, then this may motivate CISOs (and others with a vested interest in cyber security) to position the subject in this way, e.g. to explain it as a dark art. Another CISO described cyber security as ``opaque and \dots{} pervasive''; not only is it complex, it is impenetrable, and it is everywhere, implying an almost ghostly or spectral nature to the subject. 

\subsubsection{Cyber security as virtuous or moral}
Participants made reference to cyber security having dimensions of right and wrong. In particular, multiple CISOs made reference to `doing the right thing', in terms of the organisation, its employees and its third parties. Referring to the latter, CISO7 described how they are ``holding them to account to make sure they're doing the right thing''. 

Non-CISO participants also made reference to rights and wrongs. CFO2 described how breaches in cyber security can ``cause nasty accidents'' and can be ``cruel''. These value-laden references suggest the existence of the right-versus-wrong dualism. CIO1 used similar language, describing how ``nasty things happen'' in relation to cyber security. Further, CFO2 described how the CISO has ``a duty to communicate risk'', ``duty'' implying a dimension of trust and a moral aspect. They trusted the CISO to `do the right thing'.

Participants also implied aspects of cyber-security activities within their organisation as akin to social work which contributes to the conception of cyber security as virtuous; this included the CISO working with people in their organisation to ``introduce the right kind of \dots{} behaviours and judgements'' (CISO2). The notion of cyber security being interpretive is also supported by this idea of `rights and wrongs', both in terms of there being a right way to `do' cyber security -- e.g. ``you still need to get the basics right'' (CISO6) -- and a `right and wrong' dualism inherent in the topic -- e.g. ``do the right thing'' (CISO11), ``we had an army \dots{} trying to find you [employees] doing wrong'' (CISO8). In order to determine right and wrong a `moral expert' is required to interpret the situation. Positioning cyber security as moral helps to maintain the identity of the CISO role as an interpretive one. Further, requiring moral expertise also positions and reinforces cyber security as being an expert system. 

\subsubsection{Disciplinary aspects of cyber security}
A number of references were made to cyber security within organisations involving aspects of policing and enforcement. An example of this was provided by CISO5 who described how ``security's job is to hold their [the IT department's] feet to the fire''. CISOs described their activities as ``policing'' (three CISOs), discussed enforcing both penalties (three CISOs) and remedial action (two CISOs), and escalating incidents of non-compliance (two CISOs). There was an expectation of the latter from CFO2 who described how ``if there was any poor behaviour \dots{} [including from] myself, there would be an outlet for that person [the CISO] to raise the whistle''. There was also an implication of governance by consent from a number of CISOs; CISO12 described being proud of having ``change[d] the conversation and the outlook around security \dots{} from being something that is done to them to being something that is done with them''. CISO13 referred to the COVID-19 pandemic as having ``created a dynamic that is probably even less I guess accepting of governance and controls'', suggesting that governance requires consent.

These references to discipline and, in particular, punishment are predicated on a concept of right and wrong; it is difficult to punish someone if they have not done something that is considered to be wrong, if they have not transgressed against something. This further reinforces the narrative of cyber security being an expert system; because discipline is possible/permitted/accepted, then this suggests an authority of some description that sits behind that disciplinary ability. 

\subsubsection{Edifying aspects of cyber security}
Several CISOs positioned their role and its activities as improving for the organisation. For example, CISO13 explicitly noted the edifying role that they see themselves playing, stating: ``I see it on myself to ultimately get them [the organisation] to a better place and improve it''. This included positioning cyber security as enlightening. For example, CISO11 described ``dragging it [the company] out of the dark ages of security into the modern contemporary security daylight'' and the CISO6 described how, without them, the organisation ``would wander round blindly''. Multiple CISO participants made reference to changing ``mindsets''; for example, CISO4 described how they had ``changed \dots{} perception \dots{} and actually created a much more open, enabling mindset'' in relation to their stakeholders within the organisation. 

CISO8 referred to ``an altruistic element'' to cyber security, both within the organisation and in wider society, such as ``making it safe for children to use the internet \dots{} making it safe for consumers to buy bus tickets''. CISO12 stated ``I try to improve for the benefit for the many'', suggesting a self-conception as an altruist. There was also an implication that cyber security was a `cause', that CISOs believed in what they did because it was `right'. CISO8 described how they wanted a family member to ``be able to use the internet without fear''. 

Several CISOs positioned themselves as providing an ``oversight'' to the business. CISO12 described how, without them, ``the businesses [within the organisation] would suffer \dots{} because some of the things I see''. CISO2 described how, without them, their business would ``live with a kind of \dots{} fig leaf'' but eventually there would be ``a pretty serious failing''. 

\subsection{Cyber security as fearful}\label{sec:fearful}
The association of fear with cyber security was observed throughout the data. For example, words such as ``scary'' and ``worried'' were commonly used by both CISOs and non-CISOs. The perception of senior leaders in particular was that a cyber security incident was something to be afraid of, summarised by the following exchange:

\begin{quote}
\textit{Interviewer:} ``When you hear about things like the Travelex incident~\cite{tidyTravelexBeingHeld2020}, how does that tend to make you feel, how do you tend to react?''
\textit{CEO2:} ``Shit!'' [laughter]
\end{quote}

\noindent NED1 described how increased knowledge regarding cyber security, as provided by their CISO and also from what they had observed in the media, had ``got me a little scared''. CISO8 explained how ``it's very difficult for a conversation [about cyber security] not to gravitate back to being scary and inevitable''. CISOs also described deliberately utilising fear with their stakeholders, including one visceral description of the use of ``war games \dots{} with the executive committee \dots{} [whereby] you watch them shit themselves over the next half an hour'' (CISO11). CISO14 described how fear is ``coupled with `so what are you doing about it'\,'', indicating that they generate a fear response in order to position a sense of assuagement of that fear. CISO14 was also more explicit regarding assuagement, describing how they saw their role as ``giving my senior stakeholders \dots{} [a] level of comfort''. The CISO6 used similar language when referring to reporting on cyber security to their senior stakeholders, stating that ``a lot of it's about a comfort factor''. We also found that CISOs deliberately referenced cyber-security incidents affecting other organisations. As well as articulating those incidents, their communications included ``why it might not happen to us'' (CISO4).

Participants also invoked narratives of permanent emergency. For example, CISO14 described how they had ``tried to drive the message that \dots{} it's when not if [a cyber-security incident occurs]'', adding that their stakeholders ``accept that the likelihood is that we will suffer some form of cyber event over the \dots{} short to medium term''. Multiple references were made to ``sophisticated'' threats that were ``increasing'' and ``ever-changing''; CISO4 summed this up by stating that ``there are troubled times ahead''. Reference was also made to threats from ``rogue states'' and ``cyber war'', with participants referring to being ``acutely conscious of the sectors we play in, of meeting our responsibility to do our best to defend against such an attack''.\footnote{Deliberately not attributed to inhibit potential identification.}  

\subsubsection{Cybersecurity as bad news}
We found that cyber security was characterised as bad news, and the CISO as a `bad news merchant'. CISO3 described how ``our job is to point out there's a problem here'', while CISO13 referred to activities aimed at ``flush[ing] out some of the issues''. CISO2 described how the role of the CISO is ``turning over rocks and doing discovery and showing how bad everything is so things appear to get worse''. CISO12 provided an example of the CISO being a `bad news merchant', describing how they would only be present at a board meeting ``if it's something going wrong''. CEO1 described how ``if something serious happened I would be told quickly'', again suggesting that the CISO's primary engagement with senior leaders is to articulate bad news. However, CFO2 implied that they did not want to be told bad news, and CISO9 described how their CEO was, at one stage, ``very angry'' when presented with cyber-security risk.\footnote{This also demonstrates another aspect of cyber security that emerged from the data: it is often experienced emotionally. As well as fear and anger, panic, shame, and even love were articulated by participants in relation to cyber security.} CISO5 described how ``there's only so much [that] people can put up with'', which may indicate that bad news relating to cyber security results in fatigue. This statement also suggests a level of frustration and resignation on behalf of the CISO in relation to being the bad news merchant. CEO1 stated that ``what we try to avoid is shroud-waving at ourselves the whole time because people then don't take it seriously'', indicating again that cyber security has a negative association and suggesting the risk of `cyber fatigue'. 

The data also included a number of references to the high cost of cyber security, which may also contribute to a conception of the subject as bad news. For example, CISO8 described how ``the board is saying `we seem to be spending a lot of money on security'\,'' and another CISO described how, to their stakeholders, ``we just seem to be costing money''. Senior leaders also indicated concerns with cost; for example, CEO1 described how ``[security teams] are by nature quite expensive to run''. Beyond cost, CISOs also expressed concern with other labels being applied to them, including ``as someone who's being kind of too unreasonable and who's \dots{} creating too much bureaucracy, not being commercially minded'' (CISO13), which may also contribute to negative associations.

\subsubsection{Cyber security as an ontological threat}
We found that senior participants in particular considered a cyber-security incident to be an ontological threat to the organisation, affecting its ongoing viability. For example, CEO1 described cyber risk as ``potentially catastrophic'', stating that ``it could destroy the business''. Not all senior participants agreed however, with the CEO2 stating that although a cyber-security incident would be ``incredibly disruptive'' and would require ``alternative ways of operating'', they did not consider it as something that could put them out of business. CISOs also referred to the ontological threat of cyber security, including using terms such as ``catastrophe'' (CISO11), ``debilitating'' (CISO4) and ``disastrous'' (CISO9) to refer to cyber-security incidents. Several CISOs explicitly described how such an event would pose a risk to the ongoing viability of their organisation; for example, CISO5 described how ``[in] the worst case \dots{} your business is over''.

Several references were made to an increasing cyber-security threat. CISO6 described the need to be ``cognisant of the new threats or new attack types'' and CISO1 described the need to ``keep up to pace'' with such threats. The cyber-security threat landscape was positioned as ``sophisticated, [it] changes almost on a daily basis'' (CISO14). Non-CISOs also believed that cyber-security threats were increasing, including CFO1 describing cyber security as ``a continuing moving goal post''.

\subsection{Precarity of the CISO role}\label{sec:precarity}
The data indicated that CISOs considered themselves to occupy a precarious position. CISO12 summed this up by stating ``we know that it's implicit with our role, if something goes wrong \dots{} you're the guy [that gets fired]''. The importance of cyber security itself was, however, not questioned, suggesting that what was under threat was not the existence of the function or the role itself, but the incumbent's occupation of that role. For example, CISO3 described how the cyber-security function is ``going to be there for the long term that's for sure'' but they themselves were ``not under any illusions [as] to where accountability sits''. CISO2 stated that ``if we had a terrible security failing here \dots{} I wouldn’t escape the spotlight'', also describing how their relationship with their board could change if they ``drop the ball really hard''. Non-CISOs also implied a threat to the CISO's position; for example, NED1 described the role that their board had with regard to cyber security as including the power to ``hire [and] fire people'' and their need to determine whether ``whoever is accountable for it [cyber security] has the right level of expertise''. CEO1 stated: ``you have to make judgments on the fly about whether you think somebody is bullshitting or not''.

\subsubsection{Detachment and `othering' of the CISO}
Inhabiting what they see as a precarious role contributes towards feelings of detachment and vulnerability that were also observed in the data. Multiple CISOs felt somewhat detached from their organisation, seeing themselves as a gatekeeper or an overseer. In addition, CISOs expressed feelings of isolation (``it can be very lonely in security''), of being exposed (``you can never win really'') and of frustration (``I found it a really, really hard slog to get \dots{} stakeholders to just engage with security and understand the value''). Related to this, CISOs also experienced aspects of `othering' within their organisations. For example, CISO2 described how they are ``treated like a body that needs to be negotiated with \dots{} rather [than] as something intrinsic'' and another CISO described how they were viewed as being ``of a different tribe''. 

\section{Discussion}\label{sec:discussion}
In this section, we discuss how the CISO's identity as a soothsayer is constructed, as evidenced in our findings, and employing theories of identity work and ontological security to unpack this identity. The CISOs in our study, in response to feeling ontologically insecure, performed identity work aimed at creating or maintaining a protective identity that was essential or necessary. Similarly, the existence of a CISO within an organisation may represent that organisation's response to its own feelings of ontological insecurity. We develop this further below by first discussing how the soothsayer identity is constructed in Section~\ref{sec:soothsayer}, before exploring the identity work of organisations in relation to cyber security in Section~\ref{sec:org-identity-work} and the self-serving aspects of the CISO role in Section~\ref{sec:cyber-sophistry}. Finally, we set out implications for future CSCW work and call for greater attention to be paid to theoretically informed analytical approaches, thus, picking up a discussion that has surfaced at different moments in the history of CSCW, e.g.~\cite{CSCW:BDKRKY04,kockschCaringITSecurity2018}.

Soothsaying should not necessarily be viewed in a negative light. Historically, soothsayers have been ``prophet-consultant''~\cite{underbergSoothsayerDivinerOracle2017,daemmrichThemesMotifsWestern1987} and totem \cite{pettmanLookBunnyTotem2013}, as well as ``scapegoat''~\cite{underbergSoothsayerDivinerOracle2017}. They are associated with `reading' signs and interpreting information that would be unintelligible to a non-soothsayer, e.g.~animal entrails~\cite{collinsMappingEntrailsPractice2008} or patterns in the stars~\cite{rochbergHeavenlyWritingDivination2004}. As well as `reading' information and assimilating data, they make judgements and advise their stakeholders on the path to take, even including which path is permitted. Some of this judgement is based on `gut feeling' and soothsayers, as prophets, may have self interest in the realisation of their prophecies~\cite{underbergSoothsayerDivinerOracle2017}. Soothsayers occupy a position of some jeopardy; if their stakeholders are displeased with their interpretation, or if they are seen to have failed, the soothsayer may lose their job, e.g.~\cite{singhTrumpFiresDirector2020,sharwoodTrumpFiresCybersecurity,EquifaxReleasesDetails} or worse, their life e.g.~\cite{adekunleOralTraditionHistory1994}. In more modern times, nation states have employed `futurists' to predict military threats, which include those related to cyber security; those nation states may take a futurist's predictions seriously, even when based on their own works of fiction~\cite{nissenbaumAuthorWarnsMilitary2015}.

\subsection{Constructing the soothsayer identity}\label{sec:soothsayer}
The following section discusses the identity work performed by the CISO that contributed to the development of the soothsayer theme.

\subsubsection{The CISO as protector}
Participants in this study articulated narratives wherein the CISO performs a specific role of responding to, or providing protection from, threats. This is what Czarniawska describes as a ``\textit{modern story}', in which ``society and nature cause disequilibria, and science restores equilibrium''~\cite[p. 86]{czarniawskaNarrativesSocialScience2004} (italics in original). These stories, as expressed by both CISOs and non-CISOs, position the CISO and their discipline as the ``science [that] restores equilibrium''~\cite[p. 86]{czarniawskaNarrativesSocialScience2004}.

Soothsayers or oracles were historically consulted for motives of politics~\cite{marchais-roubelatDelphiMethodRitual2011} and warfare~\cite{connorEarlyGreekLand1988}. Tropes of warfare were used by participants in this study, including references to militaristic attack-defence concepts in relation to cyber security which are common in the cyber-security industry, e.g.~\cite{adkinsRedTeamingRed2013,diogenesCybersecurityAttackDefense2018}. Such narratives are also seen in mass media, e.g.~\cite{coreraNHSCyberattackCame2017,TrumpAdministrationEscalates2019}, some with governmental origins, e.g.~\cite{bondBritainPreparingLaunch2018}, and in academia, e.g.~\cite{limnellCyberArmsRace2016,kanniainenCyberTechnologyArms2018}. In particular, the existence of a permanent emergency that necessitates specific attention being paid to cyber security appears consistent in security discourse. Through the existence of a CISO, and statements that make its presence public, a wider narrative of cyber security as permanent emergency is supported, specifically as something that is a threat which needs to be addressed by businesses. Indeed, we found that organisations considered cyber security to be an ontological threat. In order to be seen as defending the organisation from that threat, it is helpful to make that organisation \textit{feel} like it is under attack. Using militaristic language enables the CISO to maintain this feeling but, crucially, as we discuss below, the nature of the attack is positioned as mystical or specialist, requiring a certain capability in order to be successfully defended against. This enables the CISO to position themselves as a soothsayer; if there is war, then an advantage would be gained by going `into battle' with a soothsayer on one's side, particularly to advise on aspects of that war that are not well understood.\footnote{But also to act as a totem, indicating to anyone observing the warring party that it is defended against such threats. These totemic aspects are discussed further below.} This narrative therefore benefits from the existence of opaque threats that require interpretation. 

\paragraph{Masculine dimensions of cyber security.} Militaristic language has a traditionally masculine dimension and should be viewed in this light. Wider security discourse suffers from a masculine bias which perpetuates gendered language and concepts~\cite{walklateRISKCRIMINALVICTIMIZATION1997}. The cyber-security industry suffers from a lack of representation of women~\cite{peacockGenderInequalityCybersecurity2017,reed2017GlobalInformation2017,kockschCaringITSecurity2018} and cyber-security discourse, both in mass media and in academia, commonly utilises masculine tropes; for example, the positioning of cyber-security workers as ``shadow warriors''~\cite{zanuttoShadowWarriorsNo2017}. This bias contributes to the traditionally masculine and arguably paternalistic notion of being a `protector'. Notably, soothsaying historically was a profession that was dominated by male practitioners~\cite{butlerMythMagus1993}. 

\subsubsection{The CISO as interpreter of an expert system}
We found that the CISO functions as an interpreter, utilising expertise in order to enable the organisation to make decisions, similar to findings from other cyber-security researchers, e.g.~\cite{haneyItScaryIt2018}. CISOs indicated that cyber security is an expert system, comprising technical aspects, e.g. software vulnerabilities that cannot just be taken at face-value; they require interpretation in order for risks to be related to the organisation. The CISO constructs an identity as an interpreter, but also as being \textit{necessary}, with an implication that, without their role (or perhaps without them specifically), the organisation may under- or over-react to a threat. 

References that CISOs made to interpreting data and statistics relating to cyber security can also be viewed through a semiotic lens. Without the CISO, senior leaders may look at data -- i.e. signs -- and make their own, inaccurate, interpretations. Such judgement based on signs is analogous to soothsaying practices of divination, an interpretive practice performed by ``specialists''~\cite[p. 320]{collinsMappingEntrailsPractice2008}. References that senior leaders made to the capabilities of the CISO suggested something beyond knowledge, an intangible, arguably mystical, \emph{sense} for the subject, possibly akin to divination.

The existence of qualifications and professional associations regarding cyber security, e.g.~\cite{ITGovernanceIT,CybersecurityCertificationTraining}, not only support its consideration as an expert system but are also analogous to soothsaying, in which ``membership of a prophetic association''~\cite[p. 130]{bourdillonOraclesPoliticsAncient1977} provided endorsement and validation.

\subsubsection{The CISO as moral expert}
CISOs in our study positioned cyber security as having a moral dimension, something established by other cyber-security researchers, e.g.,~\cite{kockschCaringITSecurity2018}, as well as positioning their work as edifying. Viewed alongside cyber security as an expert system, this constitutes identity work that positions the CISO as a moral expert~\cite{driverMoralExpertiseJudgement2013}. Having a moral identity helps the CISO maintain their position; this may be a stronger or more stable identity than one that is solely based on a job title. A moral identity can be seen as more important, or more elevated, within a society. If so, it serves the CISO's interest to maintain a moral aspect to the subject but, as will be discussed further below, it also serves wider interests at a societal level; positioning cyber security as moral reinforces existing narratives regarding those who are `wrong', i.e. enemies. Both mass media reporting and government rhetoric in relation to cyber security is consistent in portraying both a ``growing and evolving threat''~\cite[p. 9]{Progress20162021National} and known `villains', e.g.~\cite{coreraNHSCyberattackCame2017,peelEUScramblesStop2019}. The typical geopolitical nature of those `villains' also supports Neocleous’s arguments relating security to a wider narrative with respect to the superiority of Western cultures and ideologies~\cite[9, p. 172]{neocleousCritiqueSecurity2008}. As an illustration of this, during their interview, one senior leader made reference to multiple nation states posing a cyber security threat, providing an almost `clean sweep' of the current Western view of `villains', i.e. Russia, China and North Korea, with only Iran missing. 

Indexing a wider concept of morality also serves to legitimise the domain; as morality is an established normative concept, associating cyber security with it makes the latter more acceptable, perhaps even more `real'. Operating a cyber-security function may even represent an attempt at \textit{``salvation by works''} \cite[p. 62]{baudrillardConsumerSocietyMyths1998} (italics in original) and the conception of cyber security as virtuous, and references to its practice as edifying, support such an intention.

\paragraph{Cyber security as a belief system.}
As well as moral dimensions, this study suggests that cyber security has dimensions of mysticism and interpretation, with indications of there being a doctrine, even if that doctrine is resisted. CISO participants also suggested that cyber security was, to them, a cause; that they felt a sense of duty, even a calling. They felt the need to make a difference, to improve. Further, there was an implication of being `gifted', perhaps even `chosen', exemplified in one CISO's description of a team member who they considered unsuitable to be brought into contact with senior leaders. In combination, these factors suggest that CISOs view their discipline as akin to a belief system; it has aspects of right and wrong, its practice benefits those who receive it, and it requires special skills to interpret.

Edifying aspects of cyber security included applications of `light' versus `darkness', a motif that, traditionally, has a spiritual dimension~\cite[p. 165]{daemmrichThemesMotifsWestern1987}. The ideological associations of cyber security also support the belief system analogy. Doctrinal implications made by participants included multiple references to `lay-people', as well as references to people who ``get it'' and suggestions of orthodoxy in relation to cyber security. Further, several CISOs articulated a need to `sell' the concept of cyber security to the rest of the organisation, including the promotion of positive security behaviours, a need also found by other researchers~\cite{haneyItScaryIt2018}. This may motivate not just edification, but also evangelism. We consider these implicit associations of cyber security with systems of belief as mutually reinforcing of the role that CISOs play as soothsayers.

\subsubsection{The CISO as heretic}
Being associated with a domain that has aspects of an established belief system does not preclude resistance against this system. Resistance against orthodoxy can be a response to bureaucracy in such a system~\cite[p. 104]{lambertMedievalHeresyPopular1992} and, notably, being heretical does not necessarily result in less mysticism; heresies can claim greater mysticism than the dogma they are positioned against~\cite[p. 104]{lambertMedievalHeresyPopular1992}. Being heretical supports both the conception of cyber security as an interpretive practice, and as `art not science'. Moreover, many CISOs in this study either explicitly labelled themselves as pragmatic or as not-dogmatic. The use of the word `pragmatic' (seen in 13 out of 15 CISO interviews) implies a sense of interpretation; to be ``pragmatic'' or ``realistic'', both an analysis and a judgement are required. By suggesting the existence of dogma, a practical application of knowledge can be positioned as an alternative. 

There was a strong aversion from CISOs to not being categorised as dogmatic, suggesting that there was a fear of this label being seen as the default with regard to cyber security. This is a form of identity work, an expression of being \textit{not-X}. Even though they may be heretical, they may still be `tainted' by the association with something that is seen as a dogmatic belief system and the `I'm pragmatic' protestation can be seen as an attempt to counteract that. This relates to a separate finding from the data regarding the conflicted identity of cyber-security functions. One identified conflict was whether a cyber-security function was part of an IT function. In some cases, this conflicted identity was observed regardless of reporting line. If the IT discipline is associated with dogma then performing identity work to position oneself as pragmatic can be a reaction against being seen to be part of an IT function, serving to create an identity that is not just discrete, but specifically is \textit{not-IT}. Being \textit{not-X} can be an important factor in the creation and maintenance of identity \cite{heraclidesWhatWillBecome2012} and \textit{not-IT} and \textit{not-dogmatic} may be manifestations of such identity work. 

The invocation of dogma lends support to the notion that cyber security is akin to a system of belief. Not only does it have a sense of `right and wrong', as described earlier, there is also a sense of `authority' and accepted practice. CISOs in this study resisted the call of dogma, actively positioning themselves as heretical. We speculate that heretics are more acceptable, particularly in a society (considering an organisation as a society) that is more secular, and claiming to be heretical may be a response by the CISO to being seen as an outsider.

\subsubsection{Threats to the CISO's identity}
The positioning of the CISO as a soothsayer may be performed by the incumbent as a response to perceived threats to their own identity, similar to identity issues identified by other researchers, e.g.~\cite{ashendenCISOsOrganisationalCulture2013}. Underberg describes how ``a prophet, a figure at times not at the center of social life and who sometimes delivers prophecies that will not be well received, can be in danger of becoming a scapegoat''~\cite[p. 148]{underbergSoothsayerDivinerOracle2017}, and security itself often carries associations of blame~\cite{kockschCaringITSecurity2018}. The CISO, as modern soothsayer, may lose their job as a result of proclamations which displease their masters~\cite{singhTrumpFiresDirector2020,sharwoodTrumpFiresCybersecurity} as well as for perceived failures~\cite{EquifaxReleasesDetails}. The risk of displeasing their audience may be high as ``credibility of oracles depends on their apparent wisdom, and apparent wisdom is defined by the beliefs and opinions of those to whom it must appear as wisdom''~\cite[p. 126]{bourdillonOraclesPoliticsAncient1977}. Their proclamations may also be impenetrable and ambiguous, something observed historically, e.g.~\cite{marchais-roubelatDelphiMethodRitual2011}. Perhaps being aware of the precarity of their role, CISOs are like other prophets who ``are said to know the time and place of their own deaths''~\cite[p. 152]{underbergSoothsayerDivinerOracle2017}.
 
 \paragraph{Positioning the cyber security discipline.} The positioning of cyber security as `art not science', alongside its positioning as `a dark art' may be attempts by CISOs to modify or enhance their identity in order to maintain their position. If they consider themselves to be in a precarious position, then cyber security being considered a science, i.e. as repeatable, methodical, established, is threatening, as that would imply the relative ease of their replacement. If cyber security is a `dark art', on the other hand, i.e. mystical, arcane and requiring interpretation, then their replacement is more problematic. In other words, it is easier to replace a chemist than a soothsayer. This conception as not-science and the need for a soothsayer may be facilitated or underpinned by the lack of an established scientific basis to cyber security~\cite{herleySoKScienceSecurity2017,florencioFUDPleaIntolerance2014} -- or even vice versa.

\subsection{Organisational identity work}\label{sec:org-identity-work}
The notion of the CISO as a soothsayer also supports an organisation's own identity work, particularly if the organisation experiences ontological insecurity as a result of perceived cyber-security threats, as we found in this study. The existence of a CISO is a fact often mentioned in an organisation's public statements, including annual reports and press releases; such statements can also be seen as identity work, and serve as a means to build prestige~\cite{baudrillardConsumerSocietyMyths1998} or to demonstrate ``resemblances that provide favourable social analogies''~\cite[p. 52]{douglasHowInstitutionsThink1987}. Statements made by organisations relating to cyber-security investment, including the existence of a CISO, contribute to a fetishisation~\cite[p. 5]{neocleousCritiqueSecurity2008}\footnote{Neocleous describes how security is ``fetish[ised]''~\cite[p. 5]{neocleousCritiqueSecurity2008} to drive consumption, further arguing security's embeddedness in capitalist society.} of cyber security and have a wider intent of achieving status, and legitimacy.

\subsubsection{The CISO as totem}
The CISO as soothsayer also performs a semiotic function, particularly if viewed as a form of totem; modern totems can include ``oracles, experts''~\cite{pettmanLookBunnyTotem2013}. Pettman describes how ``totemic semiotics [are used to] \dots{} frame, and make sense of, a largely hostile and uncertain world''~\cite[p. 8]{pettmanLookBunnyTotem2013}. Organisations in this study articulated considerable uncertainty in relation to cyber security, associating this with threats to the continued viability of the organisation. 

Totems provide assuagement~\cite{pettmanLookBunnyTotem2013}, a function we found to be provided by CISOs. As totem, they function as an instrument of ontological security~\cite{pettmanLookBunnyTotem2013}, and, as ``medicine-man or wizard\footnote{These gendered terms unfortunately borne out by the gender imbalance in this study, and the wider CISO community.} \dots{} ensur[e] the prosperity of the tribe''~\cite[p. 5]{butlerMythMagus1993}. As with soothsayers, totems are also used to predict the future~\cite{pettmanLookBunnyTotem2013}. The CISO as totem also serves to indicate group membership~\cite[p. 22]{pettmanLookBunnyTotem2013} and indicate to observers the moral position of the organisation~\cite[p. 7]{cormackSociologyMassCulture2004}. Referring to the existence of a CISO in an annual report can be viewed as an organisation publicly stating their membership of an implied group, i.e. organisations that are addressing cyber-security risks. Notably, these mentions are of the \textit{role} and not the \textit{named individual} who occupies the role, despite other, albeit more senior, individuals being named in such reports. Indeed, if the CISO is totemic, then the role of the CISO and its visible existence, i.e. its image, can be understood to be more important than the person who inhabits the role~\cite[p. 19]{pettmanLookBunnyTotem2013}. Being a totem contributes to the othering of CISOs; a totem may be ``a `frenemy,' who could turn on the subject at any moment''~\cite[p. 9]{pettmanLookBunnyTotem2013} and may have a disquieting effect~\cite[p. 10]{pettmanLookBunnyTotem2013}. Othering may also result from an impression that totems are ``tricksters''~\cite[p. 27]{pettmanLookBunnyTotem2013}.

\subsection{Cyber sophistry}\label{sec:cyber-sophistry}
The CISO plays a role in not just interpreting, but also relaying information. They hold a position of power, and that power provides opportunities for sophistry. A number of participants, both CISO and non-CISO, were aware of these opportunities. It is not just in the interpretation and the associated judgement but also in the communication of that judgement where the CISO can exert influence. This influence may be self-serving or may serve to support others within the organisation. Regardless, there is a suggestion from the data that their interpretations may not be entirely objective. This may result in undesirable outcomes of reduced security due to ineffective allocation of resources~\cite{florencioFUDPleaIntolerance2014}. 

As with soothsayers, CISOs can have a self-interest in the realisation of their own prophecies, a position that may motivate unscrupulous behaviour. The CISO's role in regulating information provides them with significant influence and potentially allows them to help secure their own future. Predicting an ever-increasing number of `sophisticated' threats may deter their stakeholders from considering their replacement or removal, and multiple CISO participants made reference to development of long-term cyber-security plans, perhaps indicating an attempt to secure their positions for longer. A dynamic may exist whereby the greater the perceived threat to their position, the greater the effort in securing their future. Cyber-security threats play multiple parts in this dynamic. The greater the threat, the more likely a breach is to occur that could ultimately cost the CISO their job, if they are scapegoated for it. However, the greater the threat, the more weight the CISO can put behind their own agenda. Therefore, although it may be in the CISO's interest to articulate the existence of such threat, it may be more in their interest for those threats not to exist. 

\subsubsection{Self-serving aspects of the cyber-security industry}
The wider cyber-security industry may also be motivated to perpetuate discourse that positions the discipline as requiring interpretation by experts, but also the consumption of goods. As well as benefiting the industry~\cite{neocleousCritiqueSecurity2008}, this benefits a consumption-driven society more generally~\cite{baudrillardConsumerSocietyMyths1998}. Cyber-security predictions made by other modern-day soothsayers such as so-called ``futurists'' can be seen to be influenced by purely commercial desires, e.g. \cite{nissenbaumAuthorWarnsMilitary2015}, which may motivate predictions of a worsening security climate~\cite{florencioFUDPleaIntolerance2014}. Further, narratives of permanent emergency and associated threats, as repeated by CISOs and the wider cyber-security industry, serve to maintain a position of power for that industry and also support broader societal agendas which benefit from an ongoing sense of insecurity~\cite{neocleousCritiqueSecurity2008}. 

\subsection{Implications for CSCW}\label{sec:implications}
Seen through the lenses of sociological theories of security and identity work, our research affords an account of the cyber-security function within organisations that, we argue, calls for greater attention to be paid to theoretically informed approaches in future CSCW work in this context. This would, we believe, enrich and extend research in CSCW as well as security research more widely, as we have demonstrated throughout the discussion in this section. Here, we set out some initial implications for CSCW, grounded in the findings from our work. More specifically, such implications relate to managerial and design aspects of cyber security, as well as practical implications for security practice within organisations.

\subsubsection{Managerial implications}
If we accept that cyber security is interpretive, as we have argued in this paper, and which builds on existing debates in the security research community, e.g.,~\cite{herleySoKScienceSecurity2017}, then it should not be approached as a binary, `are we secure?', determination -- it needs specialist interpreters to advise on the level of risk. This also includes accepting that the CISO is not employed to `make things secure'. Indeed, as Herly and Van Oorschot make clear, it is indisputable that ‘being secure’ cannot be proven~\cite{herleySoKScienceSecurity2017}, if indeed it can be considered as a binary state. Rather, the role of the CISO becomes one of advising on the level of risk, at least at the most senior levels of the organisation. Accordingly, rather than being a role of `securing' -- or indeed `policing' -- an organisation, the CISO role is more akin to weather forecasting. Orienting their role away from `securing' towards `weather forecasting' also serves the CISO: they may be less likely to be fired as a result of an inaccurate weather forecast (although a series of inaccurate forecasts would likely still lead to that result). Indeed, our data shows that the perceived precarity of the CISO role leads to self-serving actions. Furthermore, managers and practitioners should be conscious of the potential for cyber-sophistry and the unhelpful outcomes that can result~\cite{florencioFUDPleaIntolerance2014}. This may require additional cyber-security expertise to exist at senior levels within an organisation, possibly in a non-executive capacity, in order to identify and challenge this, particularly given the inability to disprove many security-related affirmations, as highlighted in~\cite{herleySoKScienceSecurity2017}. 

In this paper, we have unearthed and discussed cyber security as being mystical, with CISOs occupying the role of soothsayers in their organisation. This also has an impact on how we approach cyber-security education. Approaches to education relating to cyber security could either aim to demystify it, or, alternatively, to acknowledge the mysticism and thus reinforce the need for specialist interpretation. If adopting the latter approach, education of staff may be best approached as instruction in the use of systems in a secure way, with the security aspects of this education being implicit rather than explicit, and certain decisions on acceptability of risk being deferred to cyber-security specialists. For example, rather than training staff how to identify phishing emails, an organisation may place more focus on the cyber-security team filtering emails before they reach staff. Equally, rather than training developers on common cyber-security threats, organisations may focus more effort on specialist testing and associated risk assessment of systems before implementation. These options are, however, potentially problematic if they result in end users feeling less responsible for security and depending entirely on technological protections, becoming themselves powerless in the process. Additionally, end users will always need an element of preparedness for security-related incidents. However, as with much in security, educative approaches should be viewed as a continuum rather than in a binary manner, and a change in focus that results in secure behaviour without necessarily involving explicit articulation of security specifics may ultimately lead to a more user-centric security approach, as called for by others, e.g.,~\cite{reinfelderSecurityManagersAre2019}, where employee experience is foregrounded.

\subsubsection{Practical implications}
The question of whether cyber security is a dark art or not is perhaps obsolete. However, if it is \emph{considered} a dark art then systems should be designed on the basis that their security will need to be interpreted by a specialist, in order for that specialist to advise the users of that system as to the level of risk that it poses, and how it may be mitigated.\footnote{This may even go as far as advising users whether that risk is acceptable or not. Examples of this were observed in our data, where CISOs were expected to make that decision on behalf of the organisation's leaders, conflicting with normative assumptions that executive and supervisory boards alone determine risk tolerance.} To make that easier, there may need to be a minimum level of information provided with a system in order to make that task simpler or more standardised. As an analogy, often specialist parts for domestic goods have a separate `advice/information for installer' section. That is, where there is a section for the purchaser to read and familiarise themself with, there is also a separate section for the installer. Employing this analogy, we may begin to consider a similar approach for security systems: an `information for CISO' section to be provided at point of purchase. Similarly, as we noted above, developers should not approach security in a binary, 'is it secure?', manner. They may need to appreciate, particularly in organisations, that security-risk decisions require interpretation by a specialist that the organisation has appointed. 

While others have highlighted the lack of “science” in cyber security, e.g.~\cite{herleySoKScienceSecurity2017,florencioFUDPleaIntolerance2014}, and problematised its practice in organisational contexts, e.g.,~\cite{ashendenCISOsOrganisationalCulture2013,hooperEmergingRoleCISO2016,karanjaRoleChiefInformation2017}, we introduce a different perspective. Rather than attempting to make security “more scientific”~\cite[p. 114]{herleySoKScienceSecurity2017}, perhaps there is greater value in acknowledging the interpretive nature of cyber-security practice, particularly within commercial organisations, and reclaiming soothsaying as a beneficial advisory profession, rather than seeing the term in a negative light. Even sophistry can be viewed positively, with “[e]ffective communication, including pedagogy and sound argument, [being] critical to prosperity”~\cite[p. 23]{goreSophistsSophistryWealth2011}, and perhaps it is especially important to distinguish between sophistry and “rhetrickery”~\cite[p. 7]{boothRhetoricRHETORICQuest2009}. Therefore, rather than only exploring the ‘what’ of cyber-security behavioural interventions in organisations, e.g.,~\cite{kirlapposComplyDeadLong2013,nicholsonIntroducingCybersurvivalTask2018,beautementProductiveSecurityScalable2016,conwayQualitativeInvestigationBank2017,reinheimerInvestigationPhishingAwareness2020, siadatiMeasuringEffectivenessEmbedded2017}, there is value in future research also exploring the ‘how’ and identifying the most effective means of educating and communicating desirable behaviours and practices, particularly among the non-expert actors on whom effective security depends~\cite{reinfelderSecurityManagersAre2019}. Effective cyber security may depend less on policy and control, 
themselves ineffective, e.g.,~\cite{kirlapposComplyDeadLong2013,ashendenSecurityDialoguesBuilding2016,kockschCaringITSecurity2018,pollerCanSecurityBecome2017}, and more on communication and collaboration. This has an implication on recruitment. Others have already highlighted the need for CISOs to have effective communication skills, e.g.,~\cite{hooperEmergingRoleCISO2016}, and alongside this, the need for advisory and forecasting skills should also be considered. As well as being able to evaluate risk, CISOs need to be comfortable with providing advice, and making recommendations on, its acceptability, and being clear with their stakeholders as to implications of their advice. Such an approach needs to be followed with all stakeholders within the organisation, not just leaders, as collaborative engagement at all levels is crucial to effective security outcomes~\cite{ashendenCISOsOrganisationalCulture2013,reinfelderSecurityManagersAre2019}.


Organisational leaders themselves should rely less on the CISO-soothsayer as protector, totem and scapegoat and more on them as advisor, forecaster and educator. Shifts in thinking such as this will help organisations to develop and improve their security measures collectively, rather than as the responsibility of one person or function. Unclear responsibilities, and confused, misunderstood and multifarious roles, lead to conflicts within organisations that ultimately increase their cyber-security risk. The aspiration should be to build a truly collaborative approach to cyber security~\cite{CSCW:GoLuKo04}, with multiple actors playing their part~\cite{reinfelderSecurityManagersAre2019}, and resolve the otherwise disconnected and othered state of the CISO~\cite{ashendenCISOsOrganisationalCulture2013,reinfelderSecurityManagersAre2019,karanjaRoleChiefInformation2017,hooperEmergingRoleCISO2016,ashendenSecurityDialoguesBuilding2016}, as well as improving overall organisational cyber security.

\section{Conclusion}\label{sec:conclusion}
Our intention here is not to position cyber security in the same category as either astrology or haruspication, however, this study has shown that the role of the CISO comprises several functions akin to a modern-day soothsayer for an organisation. The CISO sits at a nexus within that organisation, consuming opaque information from multiple sources and making decisions based on their interpretations and what they judge to be appropriate or `the right thing' for their stakeholders. Having a soothsayer benefits the organisation, not just through the role it performs, but also how it contributes to the organisation's own identity; as a legitimate, and defended, entity. Being a soothsayer, however, has downsides. The CISO can be scapegoated or othered, and can feel detached from the rest of the organisation.

This study has implications for both those performing the CISO role and their stakeholders. For CISOs, it would be beneficial to reflect on how such an identity affects, both positively and negatively, their interactions and practice within their organisations. Such reflection can also help them to come to terms with the detachment and alienation many of them experience within those organisations. For organisational leaders, it is useful to consider whether employing a soothsayer is indeed what they have intended, and also whether this represents an abdication of responsibility in decision making. By relying on a soothsayer to `read the signs' (i.e. to provide the weather forecast) they may deliberately be facing away from confronting a problem more directly -- or perhaps they are deliberately looking for someone to blame in the event of disaster. It may indeed be beneficial for all parties if the CISO is treated as a weather forecaster instead of a security totem.

\bibliographystyle{plain}
\bibliography{local}

\end{document}